\documentclass[letter]{emulateapj}
\usepackage{graphicx}

\shorttitle{Internal dynamics and membership of NYC}
\shortauthors{Rochau et al.}

\begin{document}


\title{Internal dynamics and membership of the NGC\,3603\,Young\,Cluster from microarcsecond astrometry}


\author{Boyke Rochau\altaffilmark{1}, Wolfgang Brandner\altaffilmark{1},
  Andrea Stolte\altaffilmark{2}, Mario Gennaro\altaffilmark{1}, Dimitrios
  Gouliermis\altaffilmark{1}, Nicola Da Rio\altaffilmark{1}, Natalia
  Dzyurkevich\altaffilmark{1}, and Thomas Henning\altaffilmark{1}}


\altaffiltext{1}{Max-Planck-Institute for Astronomy, K\"onigstuhl 17, 69117 Heidelberg, Germany}
\altaffiltext{2}{I. Physikalisches Institut, Universit\"at zu K\"oln, Z\"ulpicher Stra{\ss}e 77, 50937 K\"oln, Germany}


\begin{abstract}
We have analyzed two epochs of {\it HST}/WFPC2 observations of the young Galactic
starburst cluster in NGC\,3603 with the aim to study its
internal dynamics and stellar population. Relative proper
motions measured over 10.15 yrs of more than 800 stars enable us to
distinguish  cluster members from field stars. The best-fitting isochrone
yields  $A_{\rm V}$=4.6-4.7\,mag, a distance of 6.6-6.9\,kpc, and an age of
1\,Myr for NGC\,3603\,Young\,Cluster (NYC). We identify pre-main-sequence/main-sequence
transition stars located in the short-lived radiative--convective gap, 
which in the NYC occurs in the mass range 3.5-3.8\,M$_\odot$. We also 
identify a sparse population of stars with an age of 4\,Myr, which appear to be the lower mass counterparts to
previously discovered blue supergiants located in the giant H\,{\sc II} region
NGC\,3603. For the first time, we are able to measure the internal velocity
dispersion of a starburst cluster from 234 stars with $I < 18.5$ mag to $\sigma_{\rm pm 1D}=141\pm27 \mu$as
yr$^{-1}$ ($4.5\pm0.8$\,km s$^{-1}$ at a distance of 6.75\,kpc). As
stars with masses between 1.7 and 9\,M$_\odot$ all exhibit the same velocity
dispersion, the cluster stars have not yet reached equipartition of kinetic
energy (i.e.,\ the cluster is not in virial equilibrium). The results highlight
the power of combining high-precision astrometry and photometry, and emphasize
the role of NYC as a benchmark object for testing stellar evolution
models and dynamical models for young clusters and as a template for
extragalactic starburst clusters.
\end{abstract}





\section{Introduction}
Massive young stellar clusters are outstanding objects containing copious
numbers of stars over the entire stellar mass range. With masses between
$10^4$\,M$_{\odot}$ and 10$^7$\,M$_{\odot}$ \citep{zhang,degrijs03,mengel08},
they cover the upper end of the cluster mass function and may constitute
progenitors of globular clusters
(GCs; Zhang \& Fall 1999; McCrady \& Graham 2007). While extragalactic starburst clusters,
such as those in the Antennae Galaxies, are often barely resolved
\cite{whitmore}, in Milky Way starburst clusters thousands of individual stars
can be observed. 

In addition to three clusters in the Galactic Center region (Arches,
Quintuplet, Young Nuclear Cluster), only a handful of Milky Way starburst
clusters located in spiral arms have so far been identified
\cite[e.g.][]{brandner08}. Among the spiral arm clusters, the
NGC\,3603\,Young\,Cluster (NYC), located in its namesake giant H{\sc
  ii} region NGC\,3603 \cite{kennicutt}, is the most compact and youngest
cluster with an age of $\approx1$Myr \cite{brandl,stolte04,sung} and a central
density $\rho_0\ge6\cdot10^4M_{\odot}\rm{pc}^{-3}$ \cite{harayama}. It hosts three
Wolf--Rayet stars, at least 6 O2/O3, and 30 late O-type stars \cite{moffat}, and
is extensively referenced as a template for extragalactic starburst
environments \cite[e.g.][]{lamers}.

Previously, dynamical studies of Galactic starburst clusters were largely
restricted to one-dimensional velocity dispersions derived from radial velocity
measurements of a few of the most luminous cluster members
\cite[e.g.][]{mengeltacconi-garman}. Using multi-epoch observations of GCs,
King \& Anderson (2001, 2002) and Anderson \& King (2003a)
pioneered high precision proper motion studies with the {\it Hubble Space
Telescope} ({\it HST}), which enabled the distinction of GC members from field stars,
and the study of GC dynamics and kinematics.

In 2006, we initiated an extensive observational program to obtain multi-epoch
high-angular resolution imaging observations of Galactic starburst clusters
with the aim to study their internal dynamics and global motions. Here, we
present the results of our analysis of two epochs of {\it HST} observations of NYC
separated by 10.15\,yr. Accurate proper motions enable us to get a
"clean" census of the cluster population by rejecting field stars and to
study the internal cluster dynamics. 

\begin{figure*}[t!]
\centering
\hbox{
\includegraphics[scale=0.65]{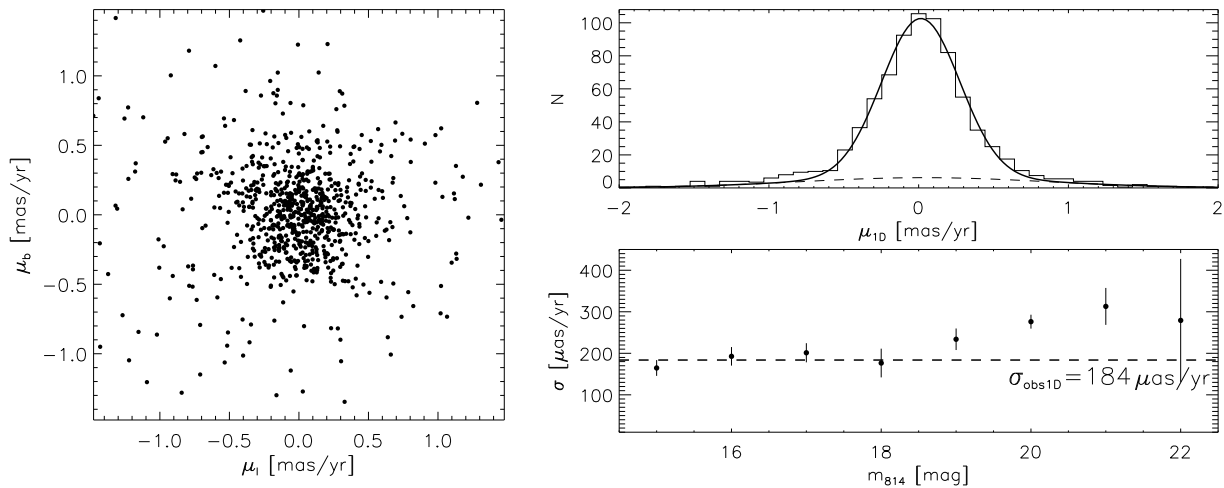}
}
\caption{\label{Fig1}{\it Left}: proper motions of stars in the NYC
  reference frame in Galactic coordinates.
 {\it Right}: histogram of the averaged one-dimensional proper motions of all stars (top),
 fitted by a two-component Gaussian (thick line), while the dashed line
 represents the distribution of the field stars. 
The lower panel depicts the observed proper motion dispersion as a function of
stellar magnitude of the cluster member candidates. 
For $14.5\,\rm{mag}<\rm{m}_{814}<18.5$\,mag, we obtain a standard
deviation $\sigma_{\rm obs 1D}=184\mu$as/yr with an uncertainty of $20\mu$as/yr (dashed line).}
\end{figure*}

\section{Observations and data analysis}\label{obs}


Two epochs of observations of NGC\,3603 with the {\it HST}/Wide-Field Planetary
Camera 2 (WFPC2) separated by 10.15 yr are analyzed. We compare Planetary
Camera (PC1) observations in F547M and F814W from epoch 1 (GO 6763, archival
data) with our second epoch observations in F555W and F814W (GO 11193). With an
image scale of 45.5\,mas $\rm{pixel}^{-1}$, PC1 provides the best point-spread 
function (PSF) sampling of the WFPC2 cameras. While the first epoch
observations were carried out in stare mode, for the second epoch we selected
a four-point sub-pixel dither pattern to facilitate bad pixel recovery and to
provide an improved PSF sampling. Table \ref{obslog} 
summarizes the observations, including individual and total exposure times.

\begin{table}
\centering
\caption{Observing log} \label{obslog}
\normalsize{
\begin{tabular}{ccccc}
\hline
Date & Filter  &   $t_{\rm exp}$(s)     & t$_{\rm tot}$(s) &  Stars \\ \hline
30/07/97 & F547M & 3$\times$1,12$\times$10,8$\times$30  &  363     &  772 \\
31/07/97 & F814W & 3$\times$0.4,12$\times$5,8$\times$20 &   221   &  1163 \\
26/09/07 & F555W & 4$\times$0.4,4$\times$26,4$\times$100 & 506      &  1048  \\
26/09/07 & F814W & 4$\times$18,4$\times$160     & 712 &   2014 \\
\hline
\hline
\end{tabular}
}
\end{table}

Data reduction has been performed with IRAF/Pyraf. We combined the individual
bias subtracted, flat-fielded images with identical exposure times using
$\texttt{multidrizzle}$ \cite{koekemoer}. This corrects for velocity
aberration and geometrical distortion including the 34th column anomaly based
on the latest distortion
correction\footnote{http://www.stsci.edu/hst/wfpc2/analysis/calfiles}
\footnote{http://ftp.stsci.edu/cdbs/uref/sad1946fu\_idc.fits}. For the
second epoch observations we applied 2$\times$2 oversampling.

Astrometry and photometry were derived from the drizzled images for each 
filter and exposure time setting using DAOPHOT
\cite{stetson} with a Penny2 PSF varying linearly across the field. Near the
faint end, where the photometric uncertainties start to increase, the star list was
filled in by the results derived from the next longer exposure, and uncertainties
assessed accordingly. The final
number of detections in each band and epoch for a $5\sigma$ threshold above
the background noise is listed in Table~\ref{obslog}. Photometric calibration
is based on the zero points provided by Dolphin (2000). Photometric 
correction for charge transfer efficiency follows the recipe provided by A.\
Dolphin\footnote{http://purcell.as.arizona.edu/wfpc2\_calib/}, and 
the astrometric correction is based on Equation 7 of 
Kozhurina-Platais et al.\ (2007) with the values for $b_1$, $b_2$, and $b_3$ 
given in Table~\ref{param} (note the different pixel scale due to 
2$\times$2 oversampling for the second epoch).

The combined image has distortions reduced to a level of 0.02 pixel 
\cite[see][]{anderson03b}. Due to the different orientation angle of 51$^\circ$ 
between the two epochs, we have to consider the uncertainty of the position induced by 
the residual geometric distortion, $\sigma_{\rm{geo}}=0.017\pm0.001\,\rm{pixel}$. 
For sinusoidal pixel phase errors \cite[see][their Figure\ 2]{anderson00}, the second epoch dithering pattern 
with 0.5 pixel shifts largely cancels out the pixel phase error. As the first epoch 
was observed in stare mode, the pixel phase error has to be considered. With a 
typical amplitude of the sinusoidal pixel phase error of $\pm$0.02 pixel, the 
uncertainty to be included in our analysis amounts to an average residual 
uncertainty of $\sigma_{\rm{pxph}}=0.013\pm0.003$\,pixel. Simulations based on TinyTim PSFs 
\citep{krist95} indicate that the positional PSF fitting uncertainty results in 
a centroiding error of $\sigma_{\rm PSF}=0.013\pm0.001$ pixel. The effect of {\it HST} breathing 
on pixel scale was determined by measuring the separation of wide pairs of stars on frames 
obtained during different phases of {\it HST}'s orbit, resulting in $\sigma_{\rm breath}=0.009\pm0.002$ 
pixel. The combined contribution of these effects amounts to 
$\sigma_{\rm{err}}=\sqrt{\sigma_{\rm{geo}}^2+\sigma_{\rm{pxph}}^2+\sigma_{\rm{PSF}}^2+\sigma_{\rm{breath}}^2}=1.21\pm0.18$\,mas.
The observed proper motion dispersion has to be corrected for $\sigma_{\rm err}$ to 
derive the intrinsic velocity dispersion of the cluster members.

\begin{table}[htb]
\centering
\caption{Parameters of the astrometric correction} \label{param}
\normalsize{
\begin{tabular}{ccccc}
\hline
Epoch & $b_1$(px)  & $b_2$(px\,$\rm{mag}^{-1}$) & $b_3$ \\ \hline
1997.58 & 0.094 & -0.0056 & $6.17\times10^{-5}$ \\
2007.73 & 0.122 & -0.011 & $3.31\times10^{-5}$ \\
\hline
\hline
\end{tabular}
}
\end{table}

As the orientation angles of the two epochs differ ($\Delta \Theta =51^\circ$),the 
common field available to our analysis is a circle with a diameter of $30\arcsec$. A 
geometric transformation based on a preliminary list of main-sequence (MS) cluster 
members is derived using IRAF/GEOMAP with a second-order polynomial. Stars detected 
in {\it I} band with photometric PSF fitting uncertainties $\sigma_{\rm{phot}} < 0.1$\,mag 
are matched after applying this transformation, and individual proper motions are 
calculated for each star. In this reference frame, cluster members center around (0,0) 
in the proper motion vector point diagram. The final proper motion table contains 829 matched stars.

\section{Proper motions and membership}\label{motions}

In Figure~\ref{Fig1} (left) we show the measured proper motions for all stars
with respect to the cluster reference frame in Galactic coordinates. 
The symmetrical distribution of all proper motions around
(0,0) indicates the absence of any large relative motion of the cluster with
respect to the field, i.e., the cluster follows the Galactic rotation
curve. As NYC's Galactic longitude of l=291.6$^\circ$ implies an
almost tangential view into the Carina spiral arm, the distributions of cluster
member and field star proper motions are superimposed.
This proper motion distribution 
is fitted by a two-component Gaussian with the wide function describing 
predominantly foreground stars, and hence non-cluster members (dashed line in
Figure~\ref{Fig1}, top right), and the narrow component is interpreted as cluster member candidates. 
We calculate cluster membership probabilities $P_{\rm mem}$ as
described in Jones \& Walker (1988) and consider stars as cluster members if
$P_{\rm mem} > 0.9$. This significantly reduces the number of contaminating 
field stars, but due to the similar proper motions of cluster and field, 
some field stars might remain in the cluster sample. Based on Besan\c{c}on models 
\cite{robin}, we estimate the number of field stars in our field 
of view to be 46 stars between $16\,\rm{mag} < m_{555} < 25$\,mag. Including stars 
with a membership probability above 
0.9 in our cluster sample, we subtracted a total of 58 stars as field stars. 
For the variable stars {\it HST} 12, 474, 481, and 574, studied by Moffat et al.\
(2004), we found membership probabilities $P_{\rm mem}$ of 0.75, 0.98, 0.90,
and 0.98, respectively, indicating that the latter three are likely cluster
members.

\begin{figure*}[htb]
\centering
\includegraphics[height=7.25cm,width=13cm]{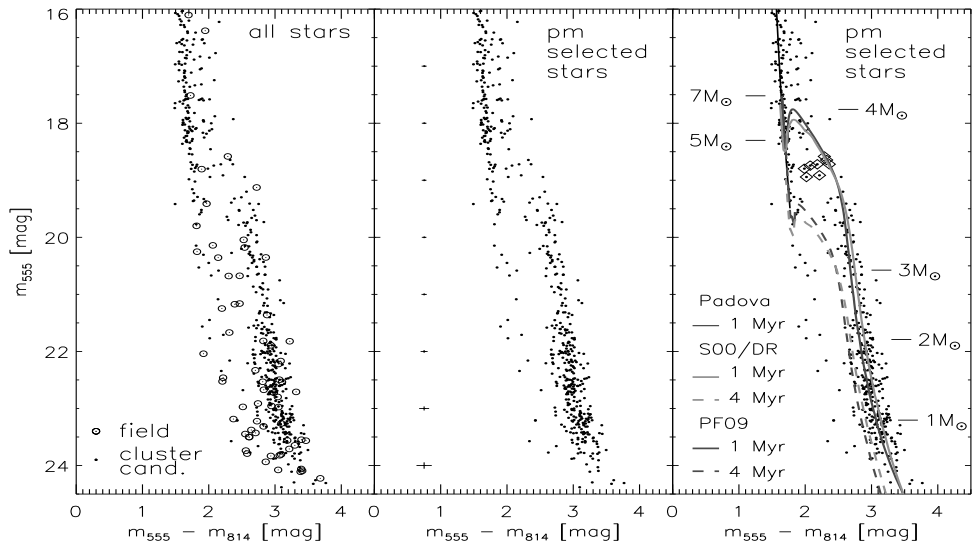}
\caption{\label{Fig2}Left: $m_{555}-m_{814}$ vs. $m_{555}$ CMD of
  cluster (small dots) and field stars (open circles). Clearly visible are the
  MS and PMS loci at $m_{555}-m_{814}\sim$1.5 and 3\,mag,
  respectively. {\it Center}: cluster member candidates with $P_{\rm
    mem}>0.9$. Apparent is the narrow PMS after the proper motion
  selection. 
Right: cluster member candidates with the best-fitting
  Padova 1\,Myr (MS, black solid line) and PF09 1\,Myr as well as S00/DR
  1\,Myr (PMS, dark and light gray solid lines, respectively) isochrones
  overplotted. Diamonds mark the stars in the r--c gap. 
  The extension of the MS below the transition region is not
  covered by the younger isochrones, but is reproduced by a 4\,Myr isochrone
  (PF09, dark gray dashed line; S00/DR, light gray dashed line).
}
\end{figure*}

\section{Extinction, distance and age}\label{cmd}

The $m_{555}$, $m_{555}-m_{814}$ CMD is shown in Figure~\ref{Fig2} (left). 
The distinction between candidate cluster members (small dots) and field 
stars (open circles) as described in Section~\ref{motions} leads to a very well 
defined cluster sequence with high-mass MS stars, 
intermediate-mass stars located in the pre-main-sequence (PMS)--MS 
transition region between $m_{555}=18$\,mag and 20\,mag, and more than 300 lower 
mass PMS stars down to $m_{555} = 24.5$\,mag (Figure~\ref{Fig2}, middle).
The efficiency of the proper motion member selection is evident in 
the left panel of Figure 2, particularly among PMS and faint stars that could 
not be distinguished from cluster members from their colors alone. The foreground 
sequence blueward of the PMS does likely not belong to the cluster, suggesting 
a residual contamination of ~18 stars with 20\,mag$<m_{555}<24$\,mag.

In the following, we assume solar metallicity \cite{melena} for the cluster and 
the relation between absolute and selective extinction from Schlegel 
et al.\ (1998). 
The upper MS is well fitted by a 1\,Myr Padova isochrone \cite[black solid line along the MS in
the right panel of Figure~\ref{Fig2}]{marigo} for $A_{\rm V}$=4.7\,mag and a
distance modulus of 14.1\,mag.
For the analysis of the PMS--MS transition region and the lower mass PMS  we use Siess models
\cite{siess}, transformed by Da Rio et al.\ (2009, S00/DR), as well as
PISA-FRANEC models (Degl'Innocenti et al. 2008, PF09), transformed into 
the observational plane using ATLAS 9 model atmospheres \cite{castelli}.
The best-fitting isochrones yield a distance between 6.6\,kpc (PF09) and 
6.9\,kpc (S00/DR) and a visual extinction $A_{\rm V}$=4.7\,mag (PF09) and
4.6\,mag (S00/DR), respectively for an age of 1\,Myr.
We note that the derived selective extinction is in good agreement with $E\rm{(}B-V\rm{)}$
= 1.25\,mag as reported by Sung \& Bessell (2004), though the absolute
extinction value derived by us is slightly higher due to the use of the
Schlegel et al.\ relations. The 1\,Myr PF09 isochrone represents the PMS 
best, in particular at the PMS--MS transition region. 

At an age of 1 Myr, stars with masses between 3.5 and
3.8\,M$_{\odot}$ are expected to be in the short-lived radiative--convective
(r--c) gap phase \cite{mayne}. This phase corresponds to 
the formation of a radiative core in the interior of the stars, 
due to the increasing central temperature  \cite{iben}.
We observe eight sources in the r--c\,gap at 18.5\,mag$<m_{555}<19$\,mag
(Figure~\ref{Fig2}) and $m_{555}-m_{814}\sim2.25$\,mag (shown as 
diamonds in the right panel of Figure~\ref{Fig2}).
If their PMS nature is spectroscopically confirmed, this is the first
identification of PMS stars in this interesting evolutionary stage.

A previously unreported CMD feature is the apparent extension of the MS 
toward lower masses below the PMS--MS transition region 
($m_{555}\gtrsim18.5$\,mag). Isochrone fitting to the MS turn-on yields an age of 
4\,Myr. The derived age is consistent with recent 
estimates of the age of the two blue supergiants Sher\,23 and Sher\,25 
\cite{melena}. These stars might represent an earlier epoch of star formation 
in the giant H\,{\sc II} region \cite[see also][]{sung}.

\section{Velocity dispersion and cluster dynamics}\label{dyn}

The distribution $\sigma_{\rm obs 1D}$ of the proper motions 
$\mu_{\rm obs1D}=\frac{\mu_l+\mu_b}{2}$, as shown in the upper right panel of
Figure~\ref{Fig1}, is a combination of internal velocity dispersion and
instrumental effects, resulting in $\sigma_{\rm obs 1D}=\sqrt{\sigma_{\rm
    pm1D}^2+\sigma_{\rm err}^2}$. In the lower right panel of Figure~\ref{Fig1}
we show the observed one-dimensional proper motion dispersion  
as a function of stellar magnitude in bins of 1 mag for candidate 
cluster members. The velocity dispersion is
constant for stars with 14.5\,mag $\le {\rm m}_{\rm F814W} \le$ 18.5\,mag
($\approx$1.7--9\,M$_\odot$).

Correcting the observed one-dimensional proper motion dispersion of $\sigma_{\rm obs
  1D}=184\pm20\mu$as/yr for the instrumental effects discussed in
Section\,\ref{obs} results in an intrinsic one-dimensional velocity dispersion $\sigma_{\rm
  pm 1D}=141\pm27 \mu$as yr$^{-1}$ for stars brighter than
$m_{814}\approx$18.5\,mag, assuming a negligible effect of binary orbital
motions (Girard et al. 1989). This corresponds to $\sigma_{\rm cl
  1D}=4.5\pm0.8$\,km s$^{-1}$ at a distance of 6.75\,kpc.
The constant velocity dispersion for stars in the mass range 1.7--9\,M$_\odot$ indicates a 
lack of equipartition of energy among cluster members.
This provides a strong indication that NYC is far from virial
equilibrium.

Nevertheless, an upper limit of the cluster mass can be obtained by deriving
the virial mass ${\rm M}_{\rm dyn}$ from the observed velocity dispersion
\cite{spitzer87}:

\begin{equation}
{\rm M}_{\rm dyn}=\eta \frac{r_{\rm h}\ \sigma_{\rm cl 3D}^2 }{\rm G}\,
\end{equation}

where $\eta\approx2.5$ (weakly depending on cluster density structure), $r_{\rm
  h}$ is the half-mass radius, $\sigma_{\rm cl 3D}$ is  the three-dimensional velocity
dispersion, and {\it G} is the gravitational constant.

NYC is mass segregated with its core radius increasing with
decreasing stellar mass \cite{nuernberger02}. A lower limit on $r_{\rm h}$ for
the high-mass stars as derived from {\it HST} data is comparable to the core radius
of $\approx$0.2\,pc (Stolte 2003), whereas Harayama et al.\ (2008) based on
the analysis of near-infrared adaptive optics data estimate $r_{\rm h}$ = 0.7--1.5\,pc 
for stars in the mass range 0.5--2.5\,M$_\odot$. If we assume
$r_{\rm h}$ = 0.5\,pc and a three-dimensional velocity dispersion of $\sigma_{\rm cl 3D} =
\sqrt{3} \times 4.5\pm0.8$\,km s$^{-1}$, we derive $M_{\rm
  dyn}=17600\pm3800\,\rm{M}_\odot$.

Considering that this dynamical mass estimate provides an upper limit, it is
in agreement with photometric studies of NYC, which assigned masses
to individual stars, and  estimated the total stellar mass to
$M_{cl}\approx10000-16000\,\rm{M}_{\odot}$ \cite{stolte06,harayama}.

\section{Summary and Conclusions}\label{sum}

Based on two epochs of high-accuracy astrometric {\it HST}/WFPC2 observations
separated by 10.15\,yr, relative proper motions of 829 stars were
measured. A selection of candidate cluster members with $P_{\rm mem}>0.9$
results in a clean cluster CMD. The best-fitting isochrone yields an age of
1\,Myr, a distance of 6.6--6.9\,kpc, and $A_{\rm V}$=4.6-4.7\,mag for the
PMS and intermediate-mass MS cluster members.

Stars at the location of the short-lived radiative convective gap, 
which occurs at 3.5--3.8\,M$_{\odot}$ at the age of
NYC, are identified for the first time. We find hints of a sparse
young low-mass population with an age of $\sim 4$\,Myr, which might constitute
an earlier generation of star formation in NGC\,3603, and likely represents 
the low-mass counterparts to several blue supergiants in the vicinity of NYC.

For the first time, the internal velocity dispersion of the starburst cluster
NYC could be measured. For stars with masses 1.7\,M$_\odot<$M$<9\,$M$_\odot$, 
we determine a one-dimensional velocity dispersion of $141\pm27 \mu$as yr$^{-1}$, corresponding 
to $4.5\pm0.8$\,km s$^{-1}$ at a distance of 6.75\,kpc. From the fact that the 
velocity dispersion does not vary with stellar mass in this mass range, 
we deduce that NYC has not yet reached equipartition of energy. 
This is not entirely unexpected at the young age of the cluster, since its 
crossing time is estimated to be 1.4\,Myr by N\"urnberger \& Petr-Gotzens (2002).

The same might be true for many extragalactic starburst clusters, where mass
estimates rely on the measurements of velocity dispersions. If these clusters
are also not yet in virial equilibrium, their masses might be systematically
overestimated. Thus, NYC provides an important benchmark for our
understanding of the early dynamical evolution and the long-term survival of
young, massive stellar clusters in the Milky Way and in other galaxies.

\acknowledgments

We thank Will Clarkson for many useful comments and discussions. 
We further thank the referee whose comments and suggestions helped 
in improving this Letter. We acknowledge support from the Deutsches Zentrum 
f\"ur Luft- und Raumfahrt (DLR), F\"orderkennzeichen 50 OR 0401.
A.-S. and D.-G. acknowledge the support of the German Research Foundation (DFG) through the Emmy Noether grant
STO 496/3-1 and the grant GO 1659/1-2, respectively.

\end{document}